\documentclass[12pt]{iopart}
\usepackage{graphicx}
\usepackage{iopams}
\usepackage{url}

\begin{document}

\title{Aharonov-Bohm rings with strong spin-orbit interaction: the role of sample-specific properties}

\author{F. Nichele}
\ead{fnichele@phys.ethz.ch}
\address{Solid State Physics Laboratory, ETH Z\"{u}rich - 8093 Z\"{u}rich, Switzerland}

\author{Y. Komijani}
\address{Solid State Physics Laboratory, ETH Z\"{u}rich - 8093 Z\"{u}rich, Switzerland}

\author{S. Hennel}
\address{Solid State Physics Laboratory, ETH Z\"{u}rich - 8093 Z\"{u}rich, Switzerland}

\author{C. Gerl}
\address{Universit\"{a}t Regensburg, Universit\"{a}tsstrasse 31, 93053 Regensburg, Germany}

\author{W. Wegscheider}
\address{Solid State Physics Laboratory, ETH Z\"{u}rich - 8093 Z\"{u}rich, Switzerland}

\author{D. Reuter}
\address{Angewandte Festk\"{o}rperphysik, Ruhr-Universit\"{a}t Bochum – 44780 Bochum, Germany}

\author{A. D. Wieck}
\address{Angewandte Festk\"{o}rperphysik, Ruhr-Universit\"{a}t Bochum – 44780 Bochum, Germany}

\author{T. Ihn}
\address{Solid State Physics Laboratory, ETH Z\"{u}rich - 8093 Z\"{u}rich, Switzerland}

\author{K. Ensslin}
\address{Solid State Physics Laboratory, ETH Z\"{u}rich - 8093 Z\"{u}rich, Switzerland}

\begin{abstract}
We present low-temperature transport experiments on Aharonov-Bohm (AB) rings fabricated from two-dimensional hole gases in p-type GaAs/AlGaAs heterostructures. Highly visible $h/e$ (up to $15\%$) and $h/2e$ oscillations, present for different gate voltages, prove the high quality of the fabricated devices. Like in previous work, a clear beating pattern of the $h/e$ and $h/2e$ oscillations is present in the magnetoresistance, producing split peaks in the Fourier spectrum. The magnetoresistance evolution is presented and discussed as a function of temperature and gate voltage. It is found that sample specific properties have a pronounced influence on the observed behavior. For example, the interference of different transverse modes or the interplay between $h/e$ oscillations and conductance fluctuations can produce the features mentioned above. In previous work they have occasionally been interpreted as signatures of spin-orbit interaction (SOI)-induced effects. In the light of these results, the unambiguous identification of SOI-induced phase effects in AB rings remains still an open and challenging experimental task.
\end{abstract}

\pacs{73.23.-b, 03.65.Vf, 71.70.Ej, 85.35.Ds}


\section{Introduction}
Semiconductor nanostructures implemented in two-dimensional systems with large SOI are considered as potential building blocks for the realization of quantum information processing and various spintronic devices \cite{Wolf2001}. In this framework, nanodevices implemented in p-type GaAs/AlGaAs heterostructures have been subject to intense theoretical and experimental studies concerning physical phenomena related to carrier-carrier Coulomb interaction and SOI \cite{Grbic2008,Komijani2010}.

A two-dimensional hole gas (2DHG) in the valence band of GaAs is characterized by wave functions whose symmetry is reminiscent of atomic p-orbitals. Due to the non-zero angular momentum and to the confinement in the growth direction, carriers are effectively described as spin ±3/2 particles, for which SOI corrections (cubic in the momentum) are expected to be stronger than for their electronic counterpart \cite{Winkler2003}. Furthermore, holes in GaAs have a very high effective mass, several times larger than that of the electrons in the conduction band. The smaller Fermi energy makes the carrier-carrier Coulomb interactions more pronounced, allowing the study of many-body related effects.

It is predicted that, in a ring-shaped nanostructure penetrated by a magnetic field, the strong SOI would induce, in addition to the conventional AB phase \cite{Aharonov1959}, a second geometrical phase term. This additional term, that in the adiabatic limit is commonly referred to as a Berry phase, acts on the spin part of the particle wave function \cite{Meir1989,Aronov1993,Mathur1992,Qian1994,Nitta1999,Frustaglia2004}. The interplay between AB phase and geometrical phase is expected to result in a complex beating-like behavior of the oscillatory magneto-resistance \cite{Loss1999,Engel2000,Pletyukhov2008,Stepanenko2009}.

Previous work with AB rings fabricated in materials with strong SOI, including p-type GaAs, showed such a beating pattern which was interpreted as being related to the occurrence of the predicted Berry phase \cite{Morpurgo1998,Yau2002,Yang2004,Grbic2007}. Other experiments \cite{Konig2006,Bergsten2006,Qu2011} focused on the modulation of the AB phase at zero magnetic field produced by changing the electrostatic potential in the ring and were interpreted in terms of the Aharonov-Casher effect \cite{Aharonov1984}.
In this work we present data on AB rings embedded in p-type 2DHGs. A fabrication technology slightly different from the work reported in Ref.\,\cite{Grbic2007}, allows a much larger electrical tunability. The overall quality of the fabricated rings is demonstrated by the measurements of highly visible $h/e$ and $h/2e$ oscillation at different gate voltage settings. In the magnetoresistance curves of our devices we clearly observe amplitude modulations of the AB oscillations at finite field, as well as AB phase jumps at zero field. Both features were studied with respect to gate voltage (top gate and/or in-plane gates) and temperature.
As we will show in detail, the identification and, possibly, the suppression of sample specific features in the future is crucial. Examples in this direction (but for electron systems) have been realized in the past by averaging Fourier spectra \cite{Meijer2004} or, more recently, using large arrays of rings \cite{Nagasawa2012}.

\section{Devices fabrication and experimental setup}
We have used two carbon doped heterostructures grown on [100] oriented substrate in two different MBE systems. In both cases the layer sequence is identical and the 2DHG lies 45 nm below the surface. Sample A was provided by WW and CG while sample B by AW and DR. At 75 mK sample A and sample B showed densities of $4.5\times 10^{15}~\rm{m^{-2}}$ and $3.5\times 10^{15}~\rm{m^{-2}}$ respectively and mobilities of $30~\rm{m^2V^{-1}s^{-1}}$ and $5.0~\rm{m^2V^{-1}s^{-1}}$ respectively. For both wafers the SOI strength was extracted from the Fourier transform of Shubnikov-de Haas oscillations measured in a Hall bar geometry and found to be comparable, with a Rashba parameter $\beta$ of $10\times 10^{-29}~\rm{eVm^{-3}}$ for sample A and $8.9\times 10^{-29}~\rm{eVm^{-3}}$ for sample B. The SOI energy splitting is calculated from $\Delta_{SO}=2\beta k_1^3$, where $k_1$ is the smaller Fermi wave vector of the two spin sub-bands. The splitting $\Delta_{SO}$ results in $1.13~\rm{meV}$ and $0.8~\rm{meV}$ for sample A and B respectively. The internal magnetic field can be estimated comparing the energy splitting given by SOI to that produced by a Zeeman field. The value of the g-factor in p-type 2DHGs is still under debate \cite{Syperek2007,Kugler2009}. Assuming a g-factor of $1$, we find a corresponding Zeeman splitting of $19.4~\rm{T}$ for sample A and $13.8~\rm{T}$ for sample B.

\begin{figure}
\includegraphics[scale=1]{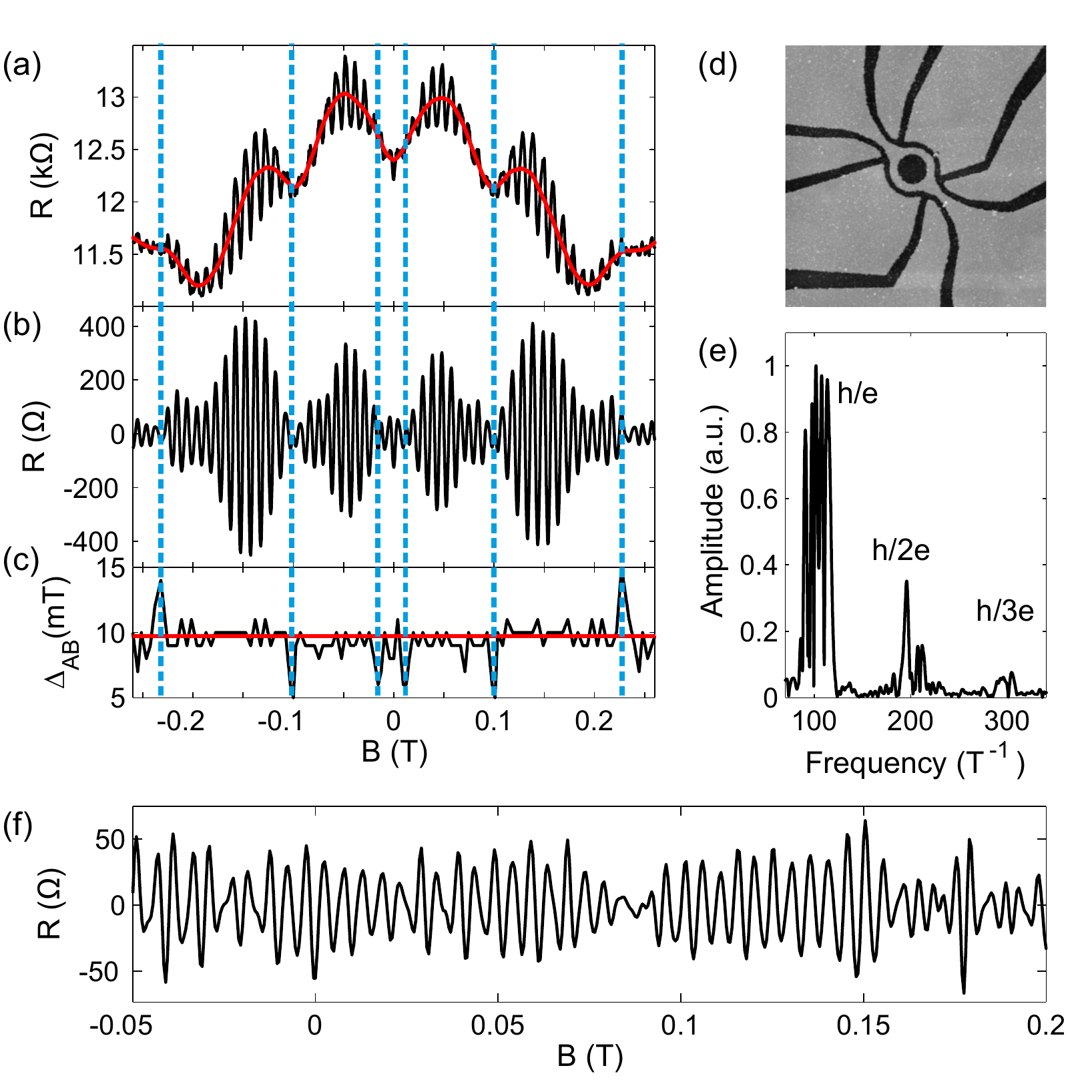}
\caption{(color online) (a) Measured four terminal magnetoresistance of the ring (black curve) together with the low frequency background which will be subtracted from the data (red curve). (b) Extracted $h/e$ oscillations. (c) Period of the $h/e$ oscillations calculated from the separation between two successive extrema. The position of the phase jumps in the $h/e$ oscillations is indicated by the blue dashed lines. (d) AFM micrograph of sample A. The black lines are 33 nm deep etched trenches, the scan frame has a lateral size of $5\times 5~{\mu\rm{m}}^2$. (e) Fourier spectrum of the raw data. (f) Extracted $h/2e$ oscillations.}
\label{fig1}
\end{figure}

Two nominally identical rings were processed with standard electron beam lithography and wet chemical etching for defining insulating trenches. An atomic force microscope scan of sample A is shown in Fig.\,\ref{fig1}(d). The dark areas indicate the etched parts, where the 2DHG underneath the surface is depleted. The area of the AFM scan is $5\times 5 ~\rm{\mu m^2}$ and the mean radius of the ring is $360~\rm{nm}$. Six in-plane gates allow tuning the transmission in the two leads and in the two arms of the ring. Further contacts are present on the same mesa on which the ring is defined (not shown in Fig.\,\ref{fig1}) to allow a four-terminal measurement of the ring resistance and the characterization of the 2DHG properties in a Hall bar geometry.
The depth of the trenches is 33 nm for sample A and 8.5 nm for sample B. In both cases the trenches are deep enough to provide confinement for holes, but the low etching depth of sample B resulted in limited electrical tunability within the leakage-free range. In order to extend its tunability, sample B was entirely covered with an insulator and a metallic top gate. The details regarding the top gate fabrication are described elsewhere \cite{Csontos2010}.
The samples were measured in a $^3\rm{He}/^4{He}$ dilution refrigerator at a base temperature of 75 mK (unless separately specified). The resistance was measured with conventional low-frequency lock-in techniques by applying a constant current of $400~\rm{pA}$ through the ring and measuring the four-terminal voltage. The low value of the current was necessary to prevent sample heating. In order to have a low resistance background, all the data discussed here have been taken with the four gates defining the transmission of the leads at the most negative possible value (just before the onset of  leakage currents).

\section{Results}
\label{results1}
For the rest of the discussion, unless explicitly stated, we will focus our attention on sample A, because it showed the highest amplitude AB oscillations. The two samples proved to be very similar, both in terms of oscillation amplitude and structure of the beatings. The electrical tuning of the ring was found to produce similar results using a combination of top gate and side gates (sample B) or just side gates (sample A). The top gate on sample B was found to have a much larger influence on the ring density than on the bulk density. This is probably due to the reduced screening in the ring as compared to the bulk and to edge effects in the ring enhancing the electric fields. In addition, the presence of the etched trenches around the ring allows an easier lateral penetration of the field lines coming from the top gate and thus leads to an improved tunability of the ring density compared to the bulk. For this reason we believe that tuning the ring with a top gate is similar to tuning it with side gates: in both cases the gates mainly change the Fermi energy and any change in the SOI has to be attributed predominantly to the change in the holes density rather than to the change of the external electric field \cite{Winkler2003, Grbic2008}.

Fig.\,\ref{fig1}(a) shows a typical magnetoresistance measurement of the ring, Fig.\,\ref{fig1}(e) shows its Fourier transform. While $h/e$ and $h/2e$ oscillations are visible in the raw data, $h/3e$ oscillations appear as a small peak in the Fourier spectrum. By filtering the Fourier spectrum of the data it is possible to separate the $h/e$ oscillations from low and high frequency components (background resistance and noise respectively). The slowly varying background that is subtracted from our data is depicted in Fig.\,\ref{fig1}(a) with a red line. The extracted $h/e$ oscillations are displayed in Fig.\,\ref{fig1}(b) while the period of the $h/e$ oscillation (calculated from the separation of successive extrema) is displayed in Fig.\,\ref{fig1}(c). We notice the presence of clear and strong beatings in the $h/e$ oscillations similar to the results reported in Ref.\,\cite{Grbic2007}. In our data the beating nodes are located at $12$ mT, $100$ mT and $230$ mT. We see in Fig.\,\ref{fig1}(c) that the $h/e$ period fluctuates around a mean value of about $10$ mT with deviations of up to $50\%$ when a beating occurs. A closer look at Fig.\,\ref{fig1}(c) shows that the period remains constant between two beating nodes and slightly changes every time a new beating occurs. 
Very similar behavior is observed for $h/2e$ oscillations as one can see in Fig.\,\ref{fig1}(f). A clear amplitude modulation is present and produces a node at $90~\rm{mT}$. Unfortunately a more quantitative analysis of the $h/2e$ oscillations is not possible, because the signal to noise ratio of the experiment does not allow us to distinguish features smaller than a few Ohms.
The beating of the magnetoconductance leads to the splitting of the $h/e$ and $h/2e$ peaks into many different sub-peaks in the Fourier spectrum shown in Fig.\,\ref{fig1}(e). In order to calculate this spectrum, we collected data in a magnetic field interval from $-650$ mT to $650$ mT, thus resolving structure in the Fourier spectrum narrower than $1~\rm{T}^{-1}$. In performing the Fourier analysis we used conventional techniques to increase the Fourier transform resolution and to suppress the borders' contribution such as removing the low-frequency background, adding zeros to both ends of the data set and windowing the data. We used different windows and different ranges of magnetic field to check that all the main features of Fig.\,\ref{fig1}(e) are genuine and not due to any finite-size effect.

The slowly varying background subtracted from the data (the red line in Fig.\,\ref{fig1}(a)) is due to the superposition of a classical magnetoresistance and aperiodic quantum conductance fluctuations widely discussed in the literature \cite{Beenakker1991}. The origin of the conductance fluctuations are spurious interference effects in the finite-size areas present along the device and their importance in this context will be addressed later on. 

\subsection{Gate dependence}
\begin{figure}
\includegraphics[scale=1]{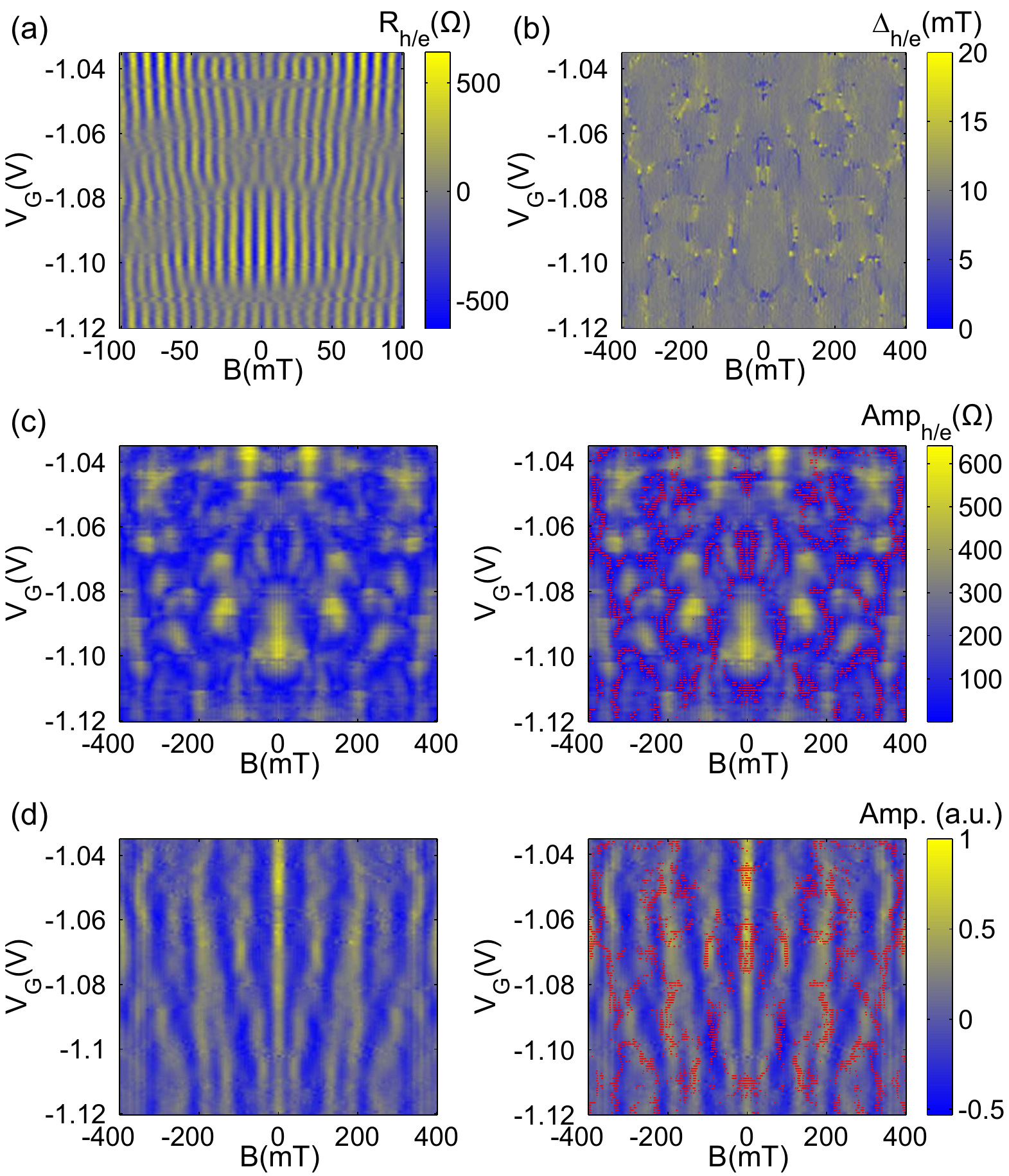}
\caption{(a) Gate dependence of the $h/e$ oscillations. (b) Calculated period of the $h/e$ oscillations in an extended field range compared to (a). The red dashed lines indicate the magnetic field range shows in (a). (c) Extracted amplitude of the $h/e$ oscillations alone (left) and compared (right) with the position of the beatings (red dots). (d) Second derivative of the slowly varying background alone (left) and compared (right) with the position of the beatings (red dots).}
\label{Gdep1}
\end{figure}

We first tuned the side gates asymmetrically to find the configuration in which the amplitude of the $h/e$ oscillations is maximized (i.e. the transmissions in the two arms are equal). Once the voltage difference that allowed having equal transmissions was found, we performed a gate dependence study by sweeping the two gates symmetrically. This is justified by the experimental observation that the two gates have very similar capacitance per unit area on the ring transmission.
In Fig.\,\ref{Gdep1}(a) we show the filtered $h/e$ oscillations as a function of the side gates voltages (symmetrically tuned): one can see that the $h/e$ oscillations are strongly affected by a change in electrostatic configuration which results in an amplitude modulation and aperiodic phase jumps of $\pi$ located at zero or finite magnetic field. In Fig.\,\ref{Gdep1}(b) we plot the $h/e$ period calculated from the same data set (but displayed for a larger interval of magnetic field values). The extracted $h/e$ period homogeneously adopts a value of $10$ mT except for particular lines where a beating occurs and the period deviates on the order of $50\%$. Comparing Fig.\,\ref{Gdep1}(a) and Fig.\,\ref{Gdep1}(b) we observe that all the phase jumps visible in Fig.\,\ref{Gdep1}(a) occur at the position of a beating of the $h/e$ oscillation.

In Fig.\,\ref{Gdep1}(c) we plot the amplitude of the  filtered $h/e$ oscillations (defined as the difference of the resistance in neighboring extrema) alone (left) and we compare it to the lines where a beating occurs (red dots superimposed to the picture on the right). We find that the two data sets are strongly anti-correlated, i.e. whenever a beating occurs, the $h/e$ oscillations experience a minimum in their amplitude.
The aperiodic conductance fluctuations mentioned before are sensitive to the gate voltage too, and result in a complex evolution of the background. This effect is visible in Fig.\,\ref{Gdep1}(d) where we plot the second derivative of the resistance background alone (left) and superimposed to the position of the beatings (right). Comparing the evolution of the background fluctuations with the beatings we observe many similarities. Although the correspondence is not perfect, in many regions the beatings are aligned with the local extrema of the background and both evolve parallel to each other along the gate voltage axis.

\begin{figure}
\includegraphics[scale=1]{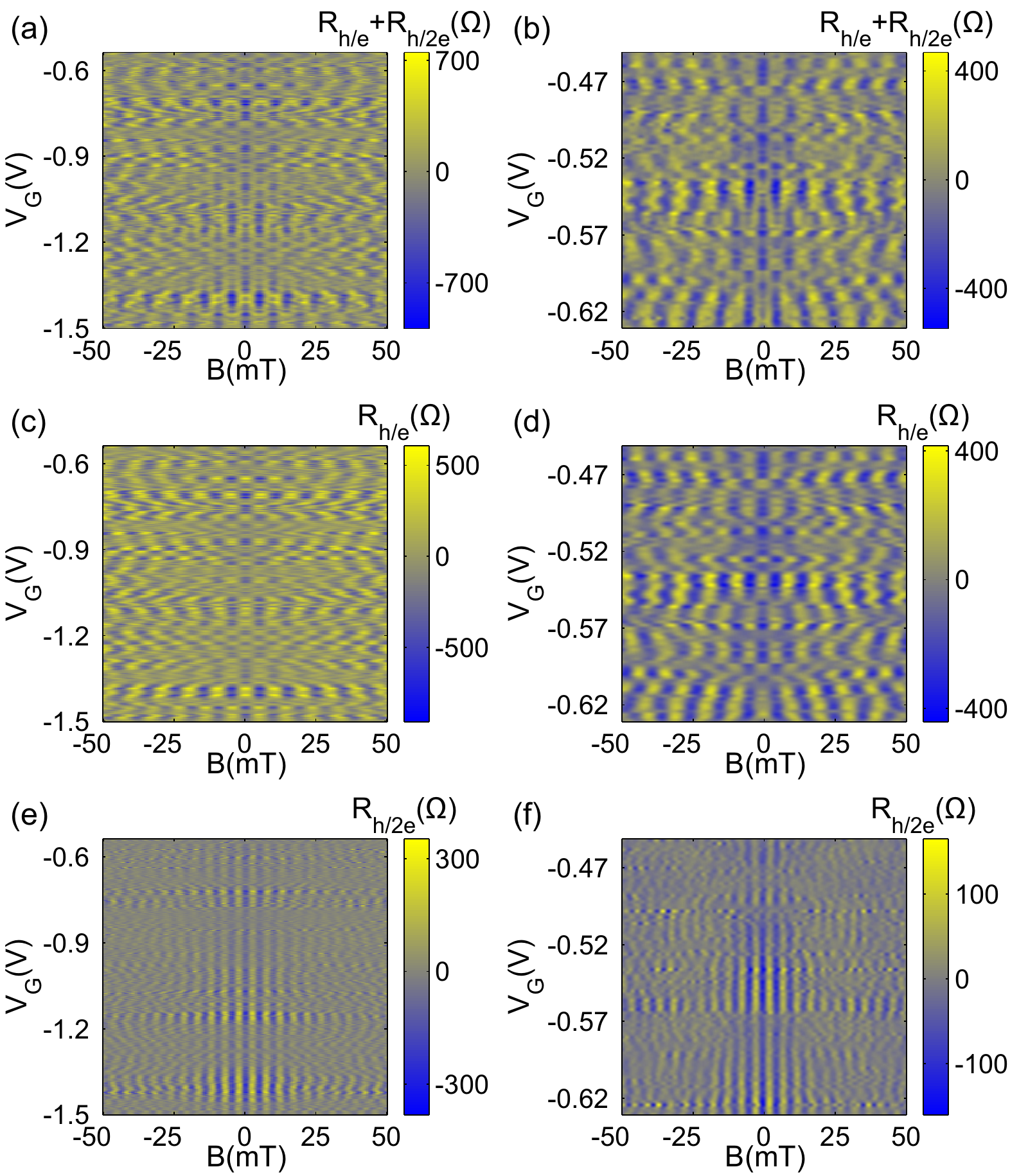}
\caption{(a) $h/e$ and $h/2e$ oscillations for sample A. (b) $h/e$ and $h/2e$ oscillations for sample B. (c) Extracted $h/e$ oscillations for sample A. (d) Extracted $h/e$ oscillations for sample B. (e) Extracted $h/2e$ oscillations for sample A. (f) Extracted $h/2e$ oscillations for sample B.}
\label{Gdep2}
\end{figure}

We also studied the behavior of our rings in a much larger range of gate voltage. The summary is shown in Fig.\,\ref{Gdep2} for both sample A (the three plots on the left, as a function of the symmetric combination of the side gate voltage) and sample B (the three plots on the right, as a function of the top gate voltage).  The superposition of $h/e$ and $h/2e$ oscillations produces a complex pattern in the magnetoresistance where certain features appear to be quasi-periodic along the voltage axis. Decomposing the different spectral contribution we can see that, even though the $h/e$ oscillations experience many phase jumps, we were never able to observe a phase jump for the $h/2e$ oscillations [see Figs.\,\ref{Gdep2}(e) and (f)].

\subsection{Temperature dependence}
Further analysis was performed by measuring the oscillations' evolution with respect to temperature. It is known that temperature affects an AB experiment in terms of phase breaking \cite{Altshuler1982} and energy averaging. Both effects eventually result in a suppression of the phase-coherent effects \cite{Hansen2001}. Increasing the temperature of the dilution refrigerator mixing chamber from $75$ mK to $300$ mK resulted in a suppression of the $h/e$ and $h/2e$ oscillation amplitude. We fitted the amplitude of the oscillations with the exponential law:
\begin{equation}
   A(T)=A(T_0)\exp{-\frac{nL}{l_{\phi}(T)}}
\label{eq_phase}
\end{equation}
where $n$ is the winding number of the oscillations, $T$ is the temperature, $l_{\phi}$ is the phase-coherence length and $L$ is the ring circumference. The effect of phase-breaking is superimposed onto the effect of energy averaging. The latter is known to introduce significant corrections in the exponential dependence of all the $h/ne$ oscillations with $n$ odd, but to leave the oscillations resulting from the interference of time-reversed paths (i.e. the $h/ne$ oscillations with $n$ even) less affected \cite{Hansen2001}. Fitting the amplitude decrease of the $h/2e$ oscillations with Eq.\,\ref{eq_phase} allows us to calculate the phase-coherence length of holes in our ring. We find $2~\rm{\mu}$m at $75$ mK, consistent with Ref.\,\cite{Grbic2007}. It should be mentioned that the coherence length calculated with this method oscillates between $1.5~\rm{\mu m}$ and $4.5~\rm{\mu m}$, depending on the magnetic field window in which the amplitude is computed. In particular, we found that close to a node in the $h/2e$ oscillations the coherence length is maximized, while close to a maximum of the envelope it is minimized (similar results are obtained for the $h/e$ oscillations). This behavior can be interpreted considering the temperature evolution of the beatings, as we will discuss later.

\begin{figure}
\includegraphics[scale=1]{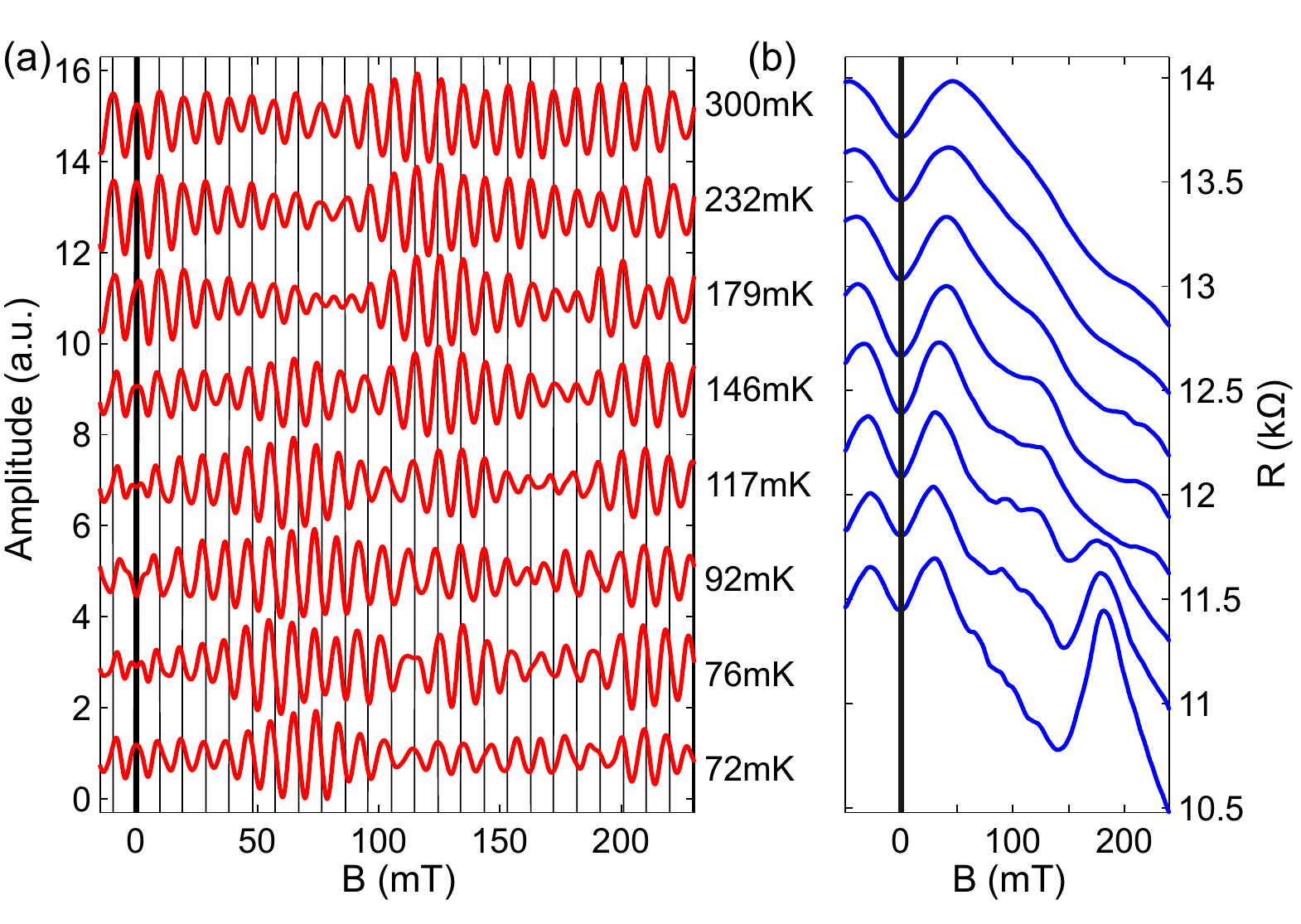}
\caption{(a) Temperature evolution of the $h/e$ oscillations superimposed to a grid with the same periodicity as the $h/e$ oscillations. Each curve has been vertically rescaled and vertically offset for clarity. The temperatures are reported on the right. The thicker vertical black line marks zero magnetic field. (b) Extracted slowly varying background for the same data set. The data are vertically offset for clarity, the vertical black line marks zero magnetic field.}
\label{Tdep}
\end{figure}

The conductance fluctuations present in the background show a strong temperature dependence of their peak amplitude, that can be fitted with an exponential law. The extracted exponents are very different for different fluctuations and gate configurations. In certain regimes the conductance fluctuations decay faster than the $h/e$ oscillations (as in Fig.\ref{Tdep}), in other cases they were still present when the $h/e$ oscillations were completely suppressed.

Fig.\,\ref{Tdep} shows a comparison of the temperature evolution of the $h/e$ oscillations (Fig.\,\ref{Tdep}(a)) and of the background (Fig.\,\ref{Tdep}(b)). All curves have been vertically shifted for clarity and the amplitude of the $h/e$ oscillations has been normalized to one. The bottom curves of Fig.\,\ref{Tdep} are taken at the lowest temperature and show similar features as those in Fig.\,\ref{fig1}. With increasing temperature we notice a gradual evolution of the beating positions along the magnetic field axis and two phase jumps at $B=0$. At the lowest temperature the oscillations have a maximum at $B=0$ that develops into a minimum at T=92 mK and again into a maximum after T=146 mK. At the highest temperature the background fluctuations are completely suppressed and the $h/e$ oscillations have a regular behavior with no phase jumps (see how the periodic grating corresponds to the oscillation maxima for every period). This behavior is not found in all the regimes investigated: in other gate configurations the conductance fluctuations decayed slower than the $h/e$ oscillations and it was not possible to observe fully suppressed beatings as in this case. We did not find any regime where beatings in the $h/e$ oscillations where still present when the background conductance fluctuations where completely suppressed. We interpret the gradual shift of the beating position, as well as the temperature dependent phase jumps at zero field, as an effect of energy averaging, as we will discuss in the next section.

\section{Analysis and discussion}
\subsection{Phase jumps and beatings}

In the framework of geometric phase effects, the beatings in the AB oscillations, as well as the splitting of the Fourier peaks, are due to the superposition of two different oscillations in the magnetoconductance of the ring. These oscillations originate from the two spin species precessing in the magnetic field texture of the ring. Berry's phase emerges in the adiabatic limit, i.e. when the electron's spin precesses many times around the effective magnetic field vector before the electron leaves the ring. In this regime, and for the case of electron rings, it was predicted that Berry's phase will make the oscillations vanish at special values of the external magnetic field, where the total magnetic field (given by the vectorial sum of the SOI induced magnetic field and the external perpendicular magnetic field) will assume ``magic'' tilt angles \cite{Malshukov1999,Engel2000}. In Ref.\,\cite{Engel2000} it is shown that an experimental observation of a splitting in the Fourier spectrum due to Berry phase would require the application of an external perpendicular field comparable to the SOI induced field. This is hard to reach in most of the experiments within the low field AB regime. In order to reach adiabaticity, the quantity $\kappa$, often referred to as adiabaticity parameter and defined as:
\begin{equation}
   \kappa=\frac{\omega_B L^2}{(2\pi)^2 D}
\label{kappa}
\end{equation}
needs to be much larger than unity. In Eq.\,\ref{kappa}, $\omega_B$ is the cyclotron frequency of the electron in a magnetic field $B$ and $D$ is the diffusion constant. To reach adiabaticity, rings with a large radius and materials with strong SOI are needed. The diffusive motion of the holes can slow down their propagation around the ring compared to ballistic motion, thus making adiabaticity easier to reach.
Using the values for our two samples we calculate that adiabaticity would be reached for a magnetic field larger than ten Tesla, where the magnetoconductance of the ring is dominated by the quantum Hall effect. The diffusion constant used in this estimate is calculated for a bulk 2DHG. In the confined geometry we are using, disorder induced scattering might be significantly enhanced making the rings adiabatic for smaller magnetic fields. Another complication might arise due to the presence of different transverse modes in the arms of the ring and their interplay with each other in determining the total conductance. From the lithographic width of the ring's arm we calculate the presence of $3-4$ transverse modes, with an energy separation of the order of $30~\rm{\mu eV}$. However the disorder potential, that we estimate as $\hbar/\tau_q$ where $\tau_q$ is the quantum scattering time, is of the order of $60~\rm{\mu eV}$, comparable to the modes' energy separation. This indicates that the modes are partially mixed, and the ring can not be considered as an ideal 1D device. It should be mentioned, however, that in the same wafer we could process, with the same fabrication technology, quantum point contacts where we could observe a well developed first quantized mode.

The presence of beatings in $h/e$ oscillations and splittings in the Fourier spectrum were often interpreted as a signature of Berry's phase as described above. However there are other phenomena, whose origin is not related to SOI, that can lead to similar effects and that were rarely taken into account. In Ref.\,\cite{Jo2007} it is shown that the interplay between different modes leads to the formation of strong beatings even in materials with low SOI. In that work it is pointed out that the beatings' evolution, if connected to mode mixing, could be affected by the electrostatic potential present in the ring and thus be modified via a gate voltage. Numerical simulations show how a phase jump of $h/e$ oscillations at $B=0$ leads to the onset of a beating that gradually evolves at finite magnetic field. The period change by $50\%$ is also presented and explained as a result of phase rigidity imposed by the Onsager relations. This effect is very similar to the measurements in our rings where, as discussed above, we estimate the presence of a few transverse modes. The strong relation between phase jumps at zero field and beatings at finite field is evident from Fig.\,\ref{Gdep1}. The small period change that occurs after a phase jump can be related to the presence of different modes with different effective radii: when the main contribution to the conductance changes from one mode to the other, also the effective area of the ring changes and the total phase might experience a phase jump of $\pi$.

In Fig.\,\ref{Gdep2} we show the gate voltage dependence in a much larger range for the two samples under study. The superposition of $h/e$ and $h/2e$ oscillations forms a complex pattern, but a line-by-line Fourier analysis shows us that only $h/e$ oscillations experience phase jumps along the gate voltage axis.
A similar pattern was observed in the work of others and interpreted as a signature of the Aharonov-Casher effect \cite{Konig2006,Qu2011}. In both Ref.\,\cite{Konig2006} and Ref.\,\cite{Qu2011} the contribution of $h/e$ and $h/2e$ oscillations was not separated. The interpretation is then not straightforward because it is not clear which features are produced by genuine phase jumps in the $h/2e$ oscillations alone and which are produced by phase jumps in the $h/2e$ oscillations or by the interplay of $h/e$ and $h/2e$ oscillations.
Assuming a ballistic single-mode 1D ring, when the density in the ring is symmetrically tuned in the two arms, $h/e$ oscillations will experience phase jumps of $\pi$ with a periodicity in $k_F$ given by $\Delta k_F L = 2\pi$. In order to convert the density change into a gate voltage change we measured the variation of the transmission of the ring with respect to the in-plane gates at high perpendicular magnetic field. The lever arm we estimate by tuning the transmission through many Landau levels is in agreement with a simple capacitor model. Assuming ideal conditions we estimate an average gate voltage spacing between successive phase jumps of $29$ mV, while the mean spacing we observed is $20$ mV. Any deviation from ideality will, in general, introduce additional phase jumps that are difficult to predict.

The situation is different for $h/2e$ oscillations: since they arise from the interference of exactly the same time-reversed path, they should not acquire any dynamical phase when traversing the ring so they are the ideal candidates to prove the relevance of a geometric phase. The existence of the Aharonov-Casher term was predicted for the case of electrons \cite{Frustaglia2004} and elegantly observed in a large array of rings in a recent experiment \cite{Nagasawa2012}. Interestingly, in the case of holes this effect should show a distinct signature, namely the frequency of the Aharonov Casher oscillations increases as a function of the spin orbit splitting \cite{Borunda2008}.
In our case we always observe a resistance minimum of $h/2e$ oscillations at $B=0$, which proves the existence of a Berry phase of $\pi$ in our rings. Apart from that, even though their amplitude shows a dependence on the gate voltage, which may have various different reasons, as one can see from Fig.\,\ref{Gdep2}, the $h/2e$ oscillations do not go through any phase jump in the voltage range accessible in our devices. The reason for the lack of phase jumps in the $h/2e$ oscillations is not clear yet. On one hand the theory of Aharonov-Casher effect in heavy holes system was always limited to ballistic single-mode rings, on the other hand also in the latter ideal situation the frequency of phase jumps in the $h/2e$ oscillations is expected to increase with increasing SOI. Having larger SOI and larger density tunability might allow the observation of beatings in $h/2e$ oscillations also in a heavy hole ring.

\subsection{Decoherence and ensemble averaging}
When the temperature of the experiment is increased two main phenomena can be observed: the suppression of phase coherent effects ($h/e$ oscillations, $h/2e$ oscillations and conductance fluctuations) and a significant modification of the beating pattern, including the phase jumps of $\pi$ at zero field. The first effect is expected from decoherence and allows to extract a coherence length for holes in our rings. The temperature dependence of the background conductance fluctuations also confirms their phase-coherent origin and thus a possible interplay with the $h/e$ oscillation phase.
We may expect that every time a conductance fluctuation completes half a period (i.e. it changes in sign), the superimposed $h/e$ oscillations will be modulated as well with a resulting phase change of $\pi$. This hypothesis is in agreement with the observation that beatings in the $h/e$ oscillations often appear to be correlated to the background conductance fluctuation, as shown in Fig.\,\ref{fig1} and Fig.\,\ref{Gdep1}. Furthermore in Fig.\,\ref{Tdep} we show that it is possible to suppress a conductance fluctuation by increasing the temperature and, once the conductance fluctuation is suppressed, the superimposed $h/e$ oscillation recovers a regular behavior. Sweeping the magnetic field we will have a superposition of different conductance fluctuations with various widths and height that will produce an aperiodic modulation of the magnetoresistance of the rings.

\begin{figure}
\includegraphics[scale=1]{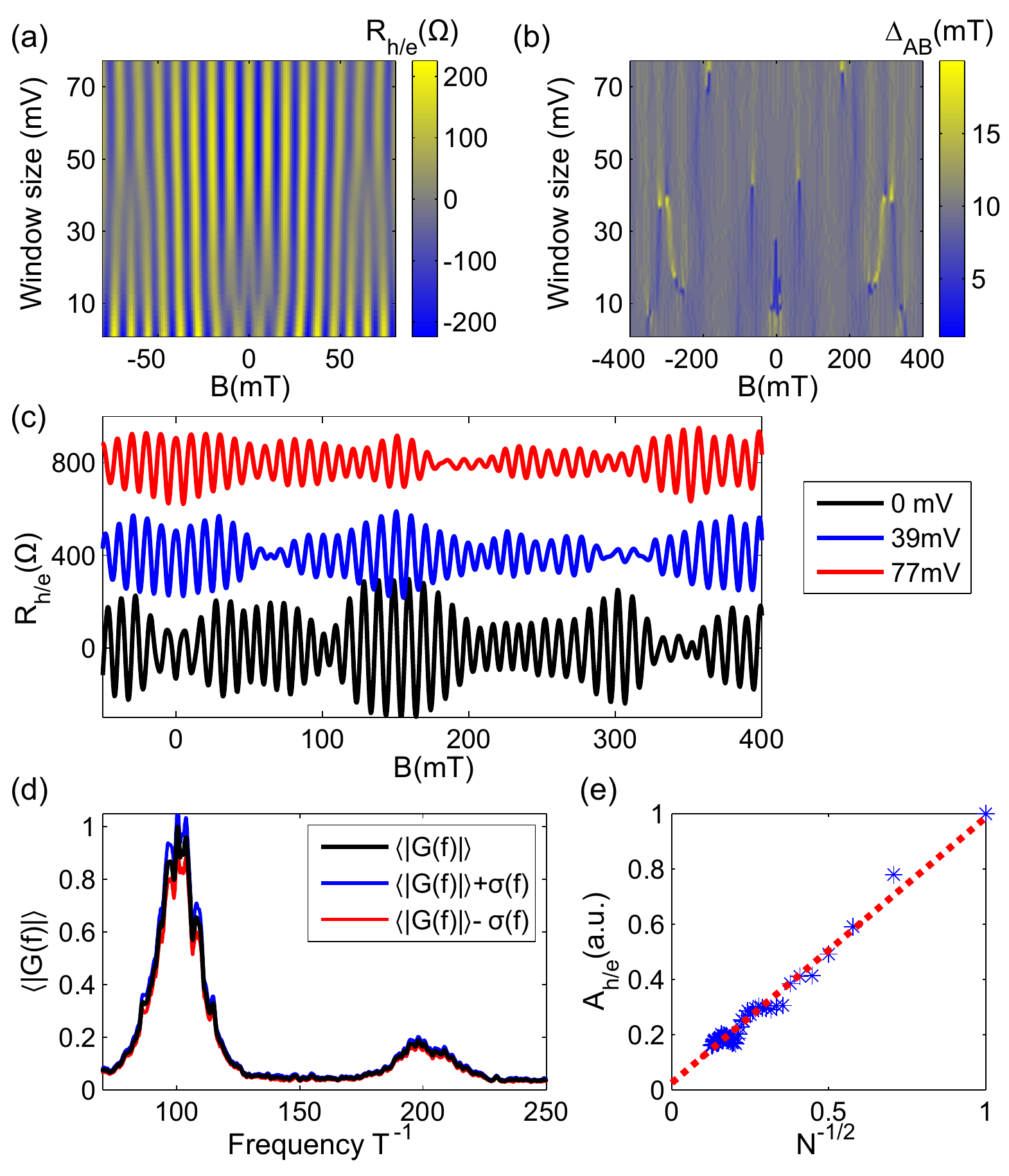}
\caption{(a) $h/e$ oscillations calculated by averaging the line at $-1.072~V$ with the nearby curves in a voltage window of increasing size. (b) Calculated period of the $h/e$ oscillations in (a). (c) Three lines extracted from (a). (d) Averaged modulus of the power spectrum for a large data set (black lines) together with the higher and lower boundary of the corresponding error bar (blue and red line respectively). (e) Amplitude of the $h/e$ oscillations as a function of the inverse square root of the number N of curves in which the average is performed (stars) together with a linear fit (dotted line).}
\label{energyaveraging}
\end{figure}

Energy averaging produces a suppression of the oscillations as well, but with a weaker temperature dependence than decoherence. The thermal length is defined by the relation $l_T=\sqrt{\hbar D/k_B T}$, and indicates the average distance after which two states separated in energy by $k_B T$ dephase by $2\pi$. The relative importance of the latter for our results can be understood considering the short thermal length of holes ($l_T=2.6~\mu m$) and comparing it with the phase-coherence length of $2~\mu m$ estimated before. In order to simulate the effect of thermal averaging we performed numerical averaging along the gate voltage axis on the same data set of Fig.\,\ref{Gdep1}. We first considered a single line taken at a gate voltage of $-1.072~V$ and then we performed numerical averaging in a gate voltage range of gradually increasing size. The final size of the averaging window we used was $73.5~\rm{mV}$, corresponding to a temperature interval of $1.5~\rm{K}$.  The results are summarized in Fig.\,\ref{energyaveraging}. In Fig.\,\ref{energyaveraging}(a) we can see the evolution of the extracted $h/e$ oscillations as a function of the magnetic field and averaging window size. One can see that the oscillations undergo various phase jumps of $\pi$ (both at zero or finite magnetic field) as the averaging window size is increased. Fig.\,\ref{energyaveraging}(b) shows the calculated $h/e$ period for the oscillations shown in Fig.\,\ref{energyaveraging}(a), where we can again observe that the beatings' position is strongly affected and shifts along the magnetic field axis as more curves are averaged. Finally Fig.\,\ref{energyaveraging}(c) shows three selected sections taken from Fig.\,\ref{energyaveraging}(a), where we compare the first not-averaged curve with the curve obtained after averaging in a gate voltage range of $39$ mV and the resulting curve obtained at the end of the averaging. One can immediately see that the amplitude of the oscillations changes by a factor of two in an interval of $1.5$ K (calculated using the gate capacitance per unit area), a very small suppression if compared to the one attributed to phase breaking. The main result is, however, the significant modification of the position of the beating nodes, visible already for a small averaging window, comparable to the temperature range used in the experiment. We can also see a phase jump in Fig.\,\ref{energyaveraging}(a) at zero magnetic field (located between $10$ and $30$ mV).

We believe these results are important regarding future analysis of SOI induced effects in AB rings since, as we show here, temperature can effectively modify the beating patterns. To the best of our knowledge the temperature dependence of the beating was never taken into account in the works regarding SOI induced effects in AB rings. An alternative way to perform energy averaging experimentally would consist in increasing the bias voltage applied to the ring. This option is not feasible in our case since Joule heating strongly suppresses the oscillations for currents larger than $1$ nA.

It was recently argued \cite{Meijer2004} that a way to suppress sample specific features in AB experiments and to observe the expected signatures of SOI induced effects is to study the average of Fourier spectra performed in an ensemble of statistical independent measurements (i.e. taken with a significant gate voltage difference). In fact, if the average is performed over the absolute value of the Fourier spectrum, it should completely suppress the phase differences of the $h/e$ oscillations present in different configurations and still not result in a cancellation of the oscillations. We checked this analysis on a data set consisting of 62 magnetoresistance traces measured by stepping the side gates voltage by 15 mV and sweeping the magnetic field from -650 mT to 650 mT. The result of the analysis is shown in Fig.\,\ref{energyaveraging}(d). We made sure that the averaged curves were statistically independent observing the decay of the $h/e$ amplitude upon successive averaging (the average of N statistically independent curves produces a $1/\sqrt{N}$ decay, as we show in Fig.\,\ref{energyaveraging}(e)). The data obtained in this way showed clearly defined $h/e$ and $h/2e$ peaks (black line in Fig.\,\ref{energyaveraging}(d)), all the small features present on the peaks are irrelevant since their size is comparable to the uncertainty given by the standard deviation (the red and blue line in Fig.\,\ref{energyaveraging}(d) represent the negative and positive limits of the error bar respectively), calculated as described in Ref.\cite{Meijer2004}. The lack of a splitting in the average of the Fourier spectra shows that the various beatings we observed as a function of gate voltage and magnetic field are features not robust against energy averaging. However caution must be paid since, as shown above, beatings of the $h/e$ oscillations can survive after averaging in a gate voltage window of moderate size.

\section{Conclusion}
We have reported measurements of large-amplitude AB oscillations in highly tunable p-type GaAs rings. In our experiments we can qualitatively reproduce various features observed in previous work (splitting of Fourier peaks, beatings, gate-dependent phase jumps) that were interpreted as signatures of SOI induced effects. Based on the discussion of the gate voltage and the temperature dependence of the features cited above, we propose an alternative origin that does not involve SOI. In particular we focus our attention on transverse mode mixing, energy averaging and interplay of the AB phase with the phase coherent conductance fluctuations. We point out that the temperature is a parameter to be taken into account, since a small energy averaging can lead to a substantial modification of the beatings. Finally we have tried to extract traces of SOI induced effects from the average of Fourier spectra taken in large ensembles of data. The results indicate that in our case most features can at least qualitatively be explained by sample-specific features.

\section{Acknowledgments}
The authors wish to thank Roland Winkler, Alex Hamilton and Alberto Morpurgo for useful discussions and the Swiss National Science Foundation for financial support.
ADW and DR acknowledge financial support from the DFG SPP1285 and the BMBF QuaHL-Rep 01 BQ 1035.

\section*{References}
\bibliographystyle{iopart-num}
\bibliography{Bibliography}

\end{document}